\documentclass[a4paper]{jpconf}
\usepackage{graphicx}
\begin{document}
\title{Muon g-2 reconstruction and analysis framework for the muon anomalous precession frequency}

\author{Kim Siang Khaw on behalf of the Muon g-2 collaboration}

\address{University of Washington, 
Department of Physics, Box 351560, Seattle, WA 98195, USA\\}

\ead{khaw84@uw.edu}

\begin{abstract}
The Muon g-2 experiment at Fermilab, with the aim to measure the muon anomalous magnetic moment to an unprecedented level of 140~ppb, has started beam and detector commissioning in Summer 2017. To deal with incoming data projected to be around tens of petabytes, a robust data reconstruction and analysis chain based on Fermilab's \textit{art} event-processing framework is developed. Herein, I report the current status of the framework, together with its novel features such as multi-threaded algorithms for online data quality monitor (DQM) and fast-turnaround operation (nearline). Performance of the framework during the commissioning run is also discussed.
\end{abstract}

\section{Introduction}

The Muon g-2 experiment~\cite{Grange:2015fou} at Fermilab aims to measure the muon anomalous magnetic moment, $a_{\mu}$ to an unprecedented precision of 140~ppb after several years of data taking. In the experiment, a longitudinally-polarized 3.1~GeV/c $\mu^{+}$ beam will be delivered by an accelerator complex (after upgrading and modifying the existing antiproton ring) at Fermilab. This beam will be injected into the muon storage ring, which is a superconducting dipole magnet, through a superconducting inflector magnet. After traversing one-quarter of the ring, the beam is ``kicked" by three pulsed magnetic kickers onto the intended storage orbit. There are four electrostatic quadrupoles installed symmetrically around the ring to focus the beam vertically. Due to the $B=1.45$~T dipole magnetic field in the ring, the muon will undergo cyclotron motion with frequency $\omega_{c}$ and the spin of the muon will precess in the magnetic field with frequency $\omega_{s}$. The difference between the two is the anomalous precession frequency, $\omega_{a}=\omega_{s}-\omega_{c}$, and it can be extracted by fitting the time modulation of the number of positrons (coming from muon decay) above an energy threshold detected by the calorimeters~\cite{Fienberg:2014kka, Kaspar:2016ofv} installed around the inside of the ring, with a fit function $N(t) = N_{0}e^{-t/\gamma\tau_{\mu}}[1+A\cos(\omega_{a}t+\phi)]$. Here, $N_{0}$ is the normalization factor, $\gamma\tau_{\mu}$ the dilated lifetime of the muon, $A$ the average asymmetry of the decay and $\phi$ the initial phase. About 400 NMR probes are installed around the ring to measure the magnetic field around the ring. The field is expressed in terms of free proton's Larmor frequency in the field after convoluted with the muon beam distribution, $\widetilde{\omega}_{p}$. $a_{\mu}$ can then be extracted using the formula $a_{\mu} = \frac{g_{e}}{2}\frac{\omega_{a}}{\widetilde{\omega}_{p}}\frac{m_{\mu}}{m_{e}}\frac{\mu_{p}}{\mu_{e}}$, where $\frac{g_{e}}{2}$ is the electron g-factor, $\frac{m_{\mu}}{m_{e}}$ the muon-to-electron mass ratio and $\frac{\mu_{p}}{\mu_{e}}$ the proton-to-electron magnetic moment ratio. This article will focus on the data reconstruction and analysis framework for the $\omega_{a}$ frequency measurement.

\section{Data processing flow}

\begin{figure}[h]
\centering
\includegraphics[width=32pc]{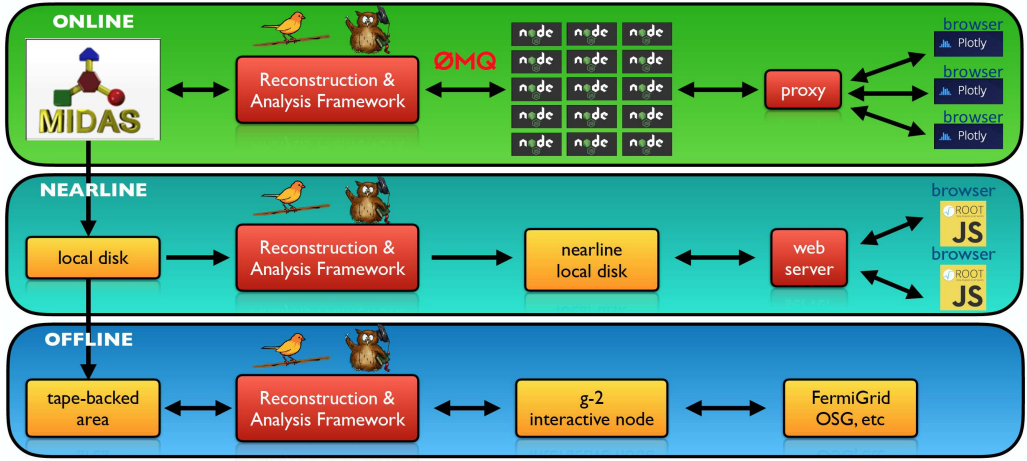}
\caption{\label{fig:dataflow}Data flow from online to nearline to offline for the Muon g-2 experiment.}
\end{figure}

The Muon g-2 event building process is done using a modern data acquisition system, MIDAS~\cite{Midas} developed at PSI and TRIUMF. Data from twenty-four calorimeters, three tracking stations, four fiber harps, Inflector Beam Monitoring System (IBMS), T-zero (T0) detector, three kickers and so on are streamed into a backend machine running MIDAS DAQ. 
It is worth noting that due to the high incoming data rate (20~GB/s), calorimeter frontend machines are equipped with NVIDIA K40 GPUs to pre-select over-threshold short pulses from 54 PbF$_{2}$ crystals per calorimeter each recorded at 800 MSPS in 700~$\mu$s long raw waveforms. The DAQ system is triggered by an FC7-based clock and control system which takes input from the Fermilab accelerator trigger system. system. The average rate at which muon bunches are injected (each bunch is defined as a \textit{fill} in subsequent paragraphs) and hence the DAQ rate is 12 Hz. The raw data is stored in the MIDAS binary format in a 40 TB RAID6 local disk before being transferred to a tape-backed area in Fermilab for permanent storage.

The data quality monitoring (DQM) system for the experiment is a custom-built framework built on top of software stacks of MIDAS, \textit{art}~\cite{Green:2012gv}, ZeroMQ~\cite{ZeroMQ}, Node.js~\cite{Nodejs} and Plotly~\cite{plotly}. Communication between MIDAS and \textit{art} is enabled using a custom-built module in \textit{art} called \textit{midastoart}. It is developed based on the MIDAS \textit{mserver} utility. 
The reconstructed and analyzed data are then aggregated using the Node servers, as shown in Fig.~\ref{fig:dataflow}(top). Finally, the data are plotted using Plotly, an online data analytics and visualization tool. A reverse proxy is also implemented to balance the query traffic.

In addition to the DQM, the experiment also has a nearline operation (see Fig.~\ref{fig:dataflow} (middle)) that provides live feedback regarding the quality of the data files and the quality of the reconstructed data.
It also produces the latest calibration data for the DQM and the offline data processing. It operates on a multi-core machine utilizing the availability of the multi-threaded modules. Reconstructed and analyzed data are stored in the \textit{art}/ROOT~\cite{Brun:1997pa} format in the nearline local disk. Histograms of high-level physics variables are also stored in separate ROOT files. Users can browse the histograms through a web server running JSROOT~\cite{Beffart:2017gxr}. 

Once the raw data files are transferred to the tape-backed area, they will be processed using a combination of FermiGrid and Open Science Grid (OSG) as shown in Fig.~\ref{fig:dataflow} (bottom). A web interface named POMS~\cite{Box:2016uzj}, which was initially developed by a Fermilab offline production team, is utilized to automate and to book-keep the production process. 

\section{Data reconstruction and analysis framework}

\begin{figure}[h]
\centering
\includegraphics[width=24pc]{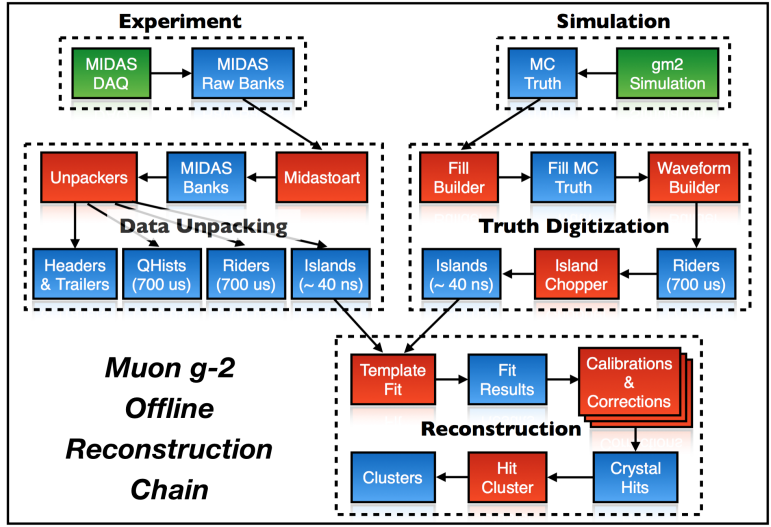}
\caption{\label{fig:offlineframework}Data reconstruction and analysis chain implemented in the \textit{art} framework for the $\omega_{a}$ measurement.}
\end{figure}

The Muon g-2 offline reconstruction and analysis chain is depicted in Fig.~\ref{fig:offlineframework}. Each stage of data processing (data unpacking, digitization, and reconstruction) consists of a series of algorithms and data products. For example, for the simulation chain, the digitization stage starts with a \textit{fill builder} which aggregate $10^{4}$ muon decay events and converts them into a single g-2 fill event. A waveform building module with the implementation of SiPM responses and digitizer behaviors is used to simulated the digitizer waveforms. Finally, an \textit{island chopping} module is used to filter out over-threshold short pulses (also called \textit{islands}) in the 700~$\mu$s long waveforms, similar to the island chopping done by the GPUs in the frontend machines. A typical reconstruction stage for the threshold-method (T-method) $\omega_{a}$ analysis starts with the pulse fitting (using a template pulse shape, $\chi^{2}$ minimization is done using Eigen linear algebra library~\cite{Eigen}) and ends with the hit clustering. Each algorithm will output a data product that serves as an input to the next algorithm. By having this modularity, it is very convenient to study alternative reconstruction algorithms developed by different collaborators.

The difference between simulated and experimental data in the reconstruction stage is that, in the simulation, all the PbF$_{2}$ crystals and the silicon photomultipliers (SiPMs) are assumed to be identical (e.g. in terms of crystal light yields and SiPM gains). Quality control of the crystals and the SiPMs have indicated that they could differ by about 20\% each regarding light transmission and breakdown voltage. Hence in practice, additional steps have to be taken to equalize them as much as possible in the experiment. A gain calibration procedure using fast pulsed laser and neutral-density filters is implemented to extract gain calibration constants (from fitted pulse integral to the number of photo-electron, \textit{npe}), and a laser monitoring system is used to monitor the SiPM gain fluctuations. Additional corrections involve monitoring the laser intensity fluctuation using pin diodes and photo-multiplier tubes (PMT). Finally the energy calibration constants (from \textit{npe} to MeV) are extracted using either the minimum ionizing particle (MIP) signals from the muons or the end-point matching of the decay energy spectrum. All these calibration and correction constants are applied to the fitted energy in sequence, and the final number is a proxy for the particle energy. On the other hand, the hit time information is extracted from the template-fitting technique, and it has a resolution of about 20~ps. A pulse-pileup separation of 100\% can be achieved for a time difference of larger than 5~ns between two pulses.

The $\omega_{a}$ offline framework also has extensive interfaces to the database of the experiment for reconstruction and analysis purposes. They are the \textit{MIDAS online database (ODB) parser} module and the \textit{database service} module. The ODB is written into the raw MIDAS file and includes most of the run information needed for data reconstruction. A dedicated ODB parser based on the boost library is implemented to extract information like the digitizer channel to crystal number mapping and the number of enabled digitizer channels. Calibration and correction constants extracted by \textit{art} modules are stored in a FNAL central database for g-2. A dedicated database service that can talk to both databases is implemented using libraries like \textit{pqxx} (Postgres C++ API)~\cite{pqxx} and \textit{boost::asio} (URL request library)~\cite{Torjo13}.

In addition to the feature mentioned above, we have also utilized the Intel Threading Building Blocks (TBB) to parallelize data processing within an art module. This parallelization is possible because our data products are in a nested structure and data from each calorimeter is independent from each other. We have implemented such parallelization in the unpacking and the reconstruction stages. A fill event processing rate of $8-10$ Hz is achievable using a 16-core CPU machine, and such a speed up enabled us to run DQM and nearline operation without missing many events.

\section{Performance during the Summer 2017 commissioning run}

During the 5-week commissioning run in Summer 2017, the event rate was about 0.15~Hz. The DAQ uptime was about 60\% and the online DQM and nearline uptimes were close to 100\%. Around 20~TB of data was recorded. The nearline data processing rate was limited by the ROOT I/O. To catch up with the DAQ rate, the ROOT compression level had to be reduced.

\begin{figure}[h]
\includegraphics[width=21pc]{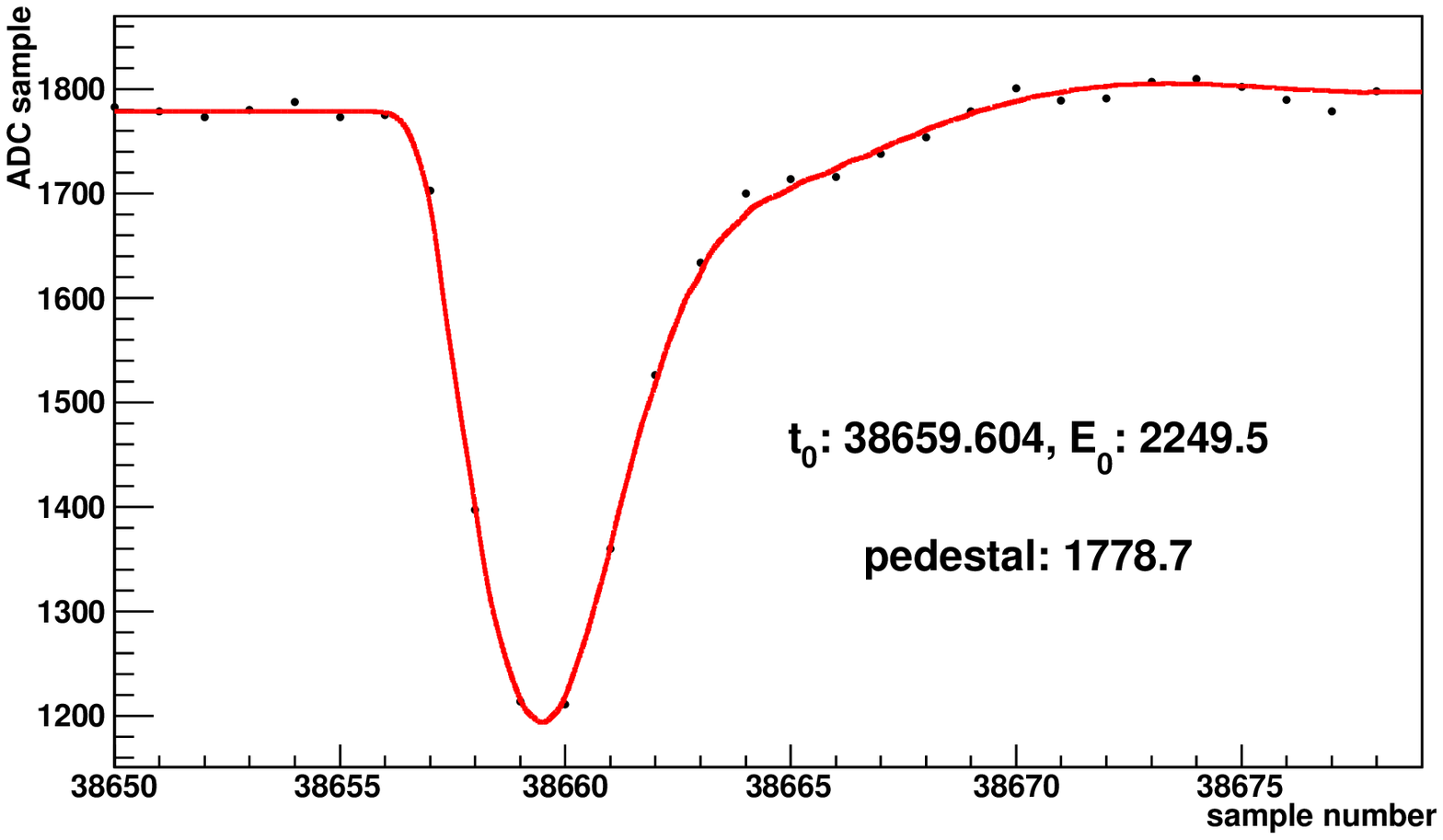}
\includegraphics[width=18pc]{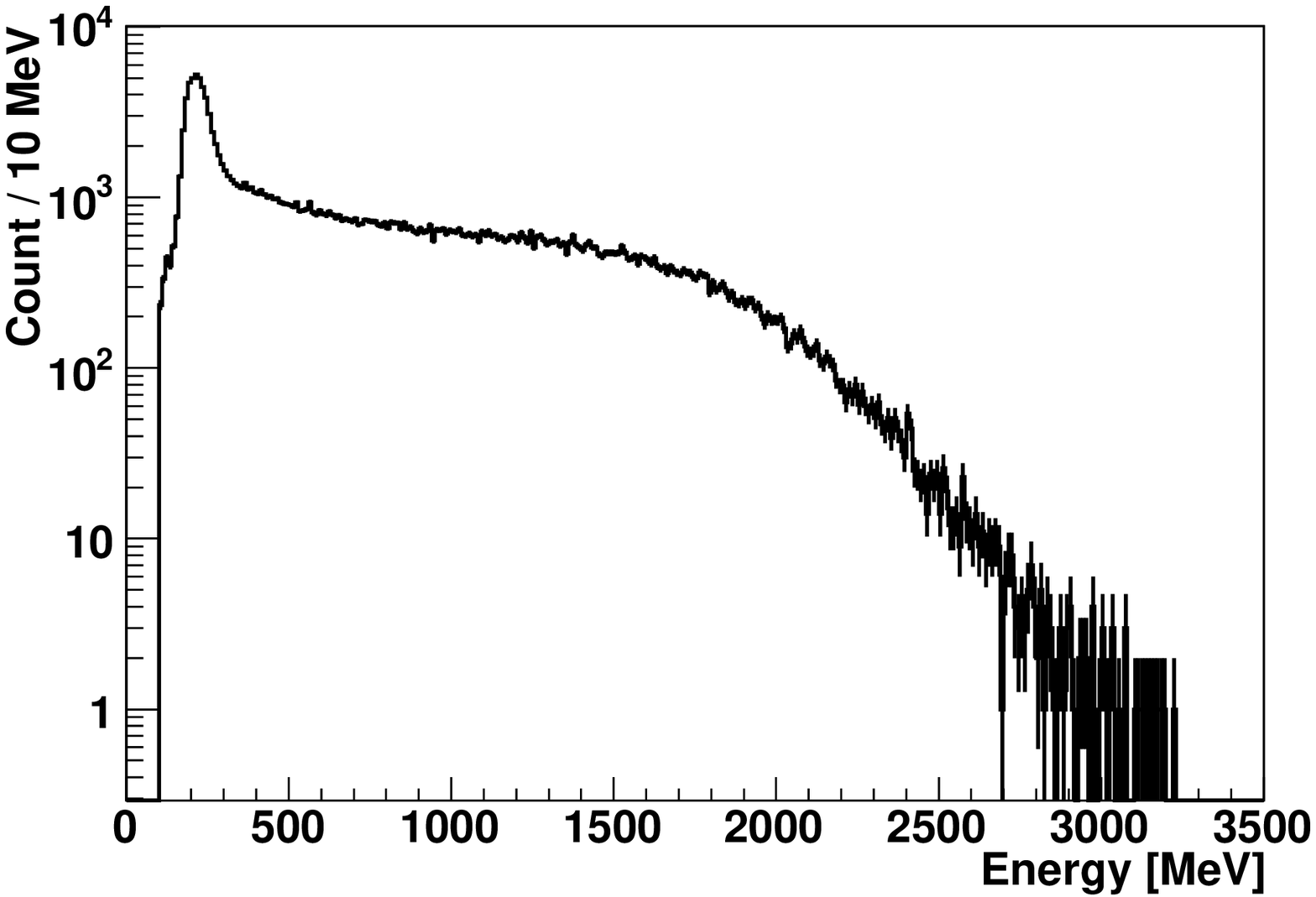}
\caption{\label{fig:recon}(left) Template fit of a digitized SiPM pulse. Red curve is the template pulse shape and black dots are the digitized samples for a typical positron impact. (right) Distribution of the reconstructed particle energy of a typical run.}
\end{figure}

\begin{figure}[h]
\includegraphics[width=20pc]{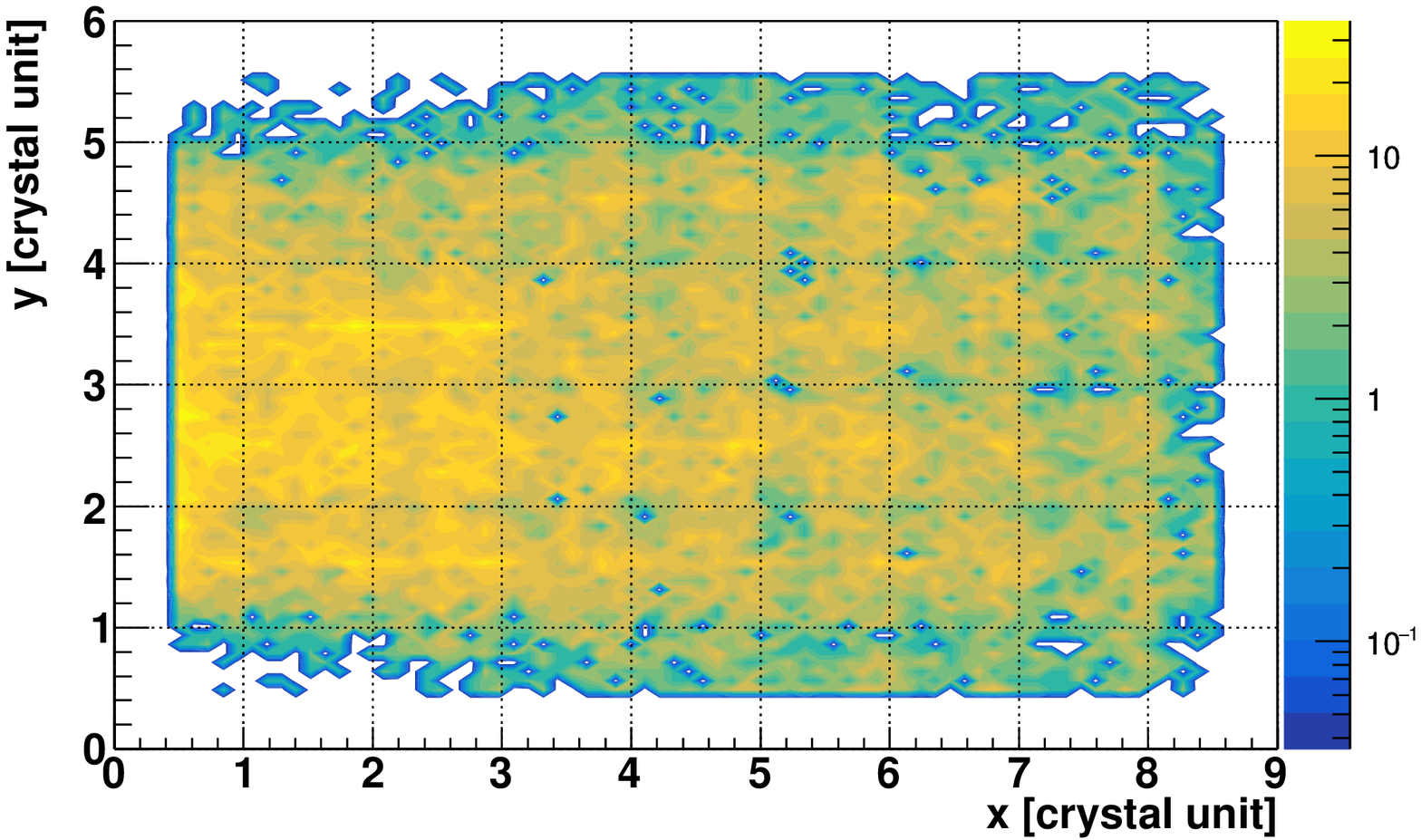}
\includegraphics[width=19pc]{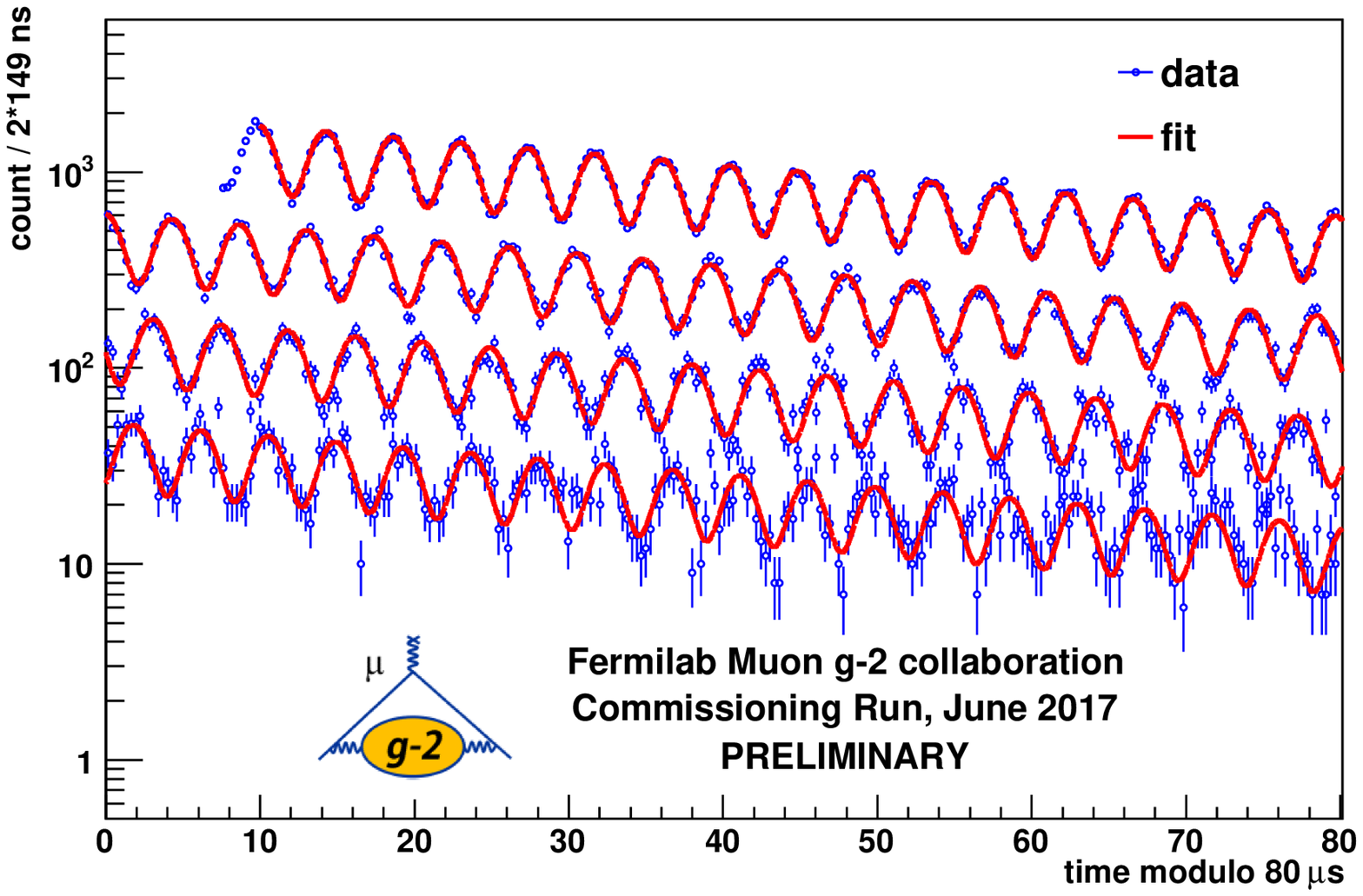}
\caption{\label{fig:wiggle}(left) Distribution of the reconstructed hit positions on calorimeters. (right) The number of positron with $E > 1.86$~GeV as a function of time detected by the calorimeters. The modulation is at the muon anomalous precession frequency.}
\end{figure}

As part of the upstream beamline that eliminates the proton and the pion beam was not ready during the commissioning run, the incoming beam was not purely muon but a mixture of the three species ($p$, $\pi^{+}, \mu^{+}$). An example of the fitted pulse is shown in Fig.~\ref{fig:recon}(left). A typical energy distribution of the detected particles is displayed in Fig.~\ref{fig:recon}(right). The low energy peak ($\approx$ 200~MeV) is dominated by the $p/\pi^{+}/\mu^{+}$ MIP signals whereas the higher energy part is coming from mainly muon-decay positrons. A typical hit distribution is shown in Fig.~\ref{fig:wiggle}(left). It is in good agreement with the simulation~\cite{gm2ringsim} where most of the hits are expected to be concentrating on the left-hand side of the calorimeter which is closer to the storage region of the beam. A preliminary result showing the modulation of the high energy positron rate is depicted in Fig.~\ref{fig:wiggle}(right).

\section{Conclusions}

In summary, the reconstruction and analysis framework for the $\omega_{a}$ analysis of Muon g-2 experiment is implemented into a highly-integrated system from online to offline using modern standardized tools. The raw detector data was first processed using GPU farms to reduce the incoming rate by a factor of 100, and the reconstruction and analysis of the data are performed using the algorithms implemented in the \textit{art} framework. The reconstruction stage is a chain of modules performing pulse fitting, energy calibration and hit clustering. The system performed very well during the Summer 2017 commissioning run and is currently being optimized based on the data collected during the commissioning run.

\section{Acknowledgments}

The author thanks A. Fienberg for useful discussions in preparing this manuscript. This work was supported, in part, by the U.S. Department of Energy Office of Science, Office of Nuclear Physics under award number DE-FG02-97ER41020. This manuscript has been authored by Fermi Research Alliance, LLC under Contract No. DE-AC02-07CH11359 with the U.S. Department of Energy, Office of Science, Office of High Energy Physics.


\section*{References}
\begin{thebibliography}{9}

\bibitem{Grange:2015fou} 
  J.~Grange {\it et al.} [Muon g-2 Collaboration],
  Muon (g-2) Technical Design Report,''
  arXiv:1501.06858 [physics.ins-det].

\bibitem{Fienberg:2014kka} 
  A.~T.~Fienberg {\it et al.},
  ``Studies of an array of PbF$_2$ Cherenkov crystals with large-area SiPM readout,''
  Nucl.\ Instrum.\ Meth.\ A {\bf 783}, 12 (2015)
  doi:10.1016/j.nima.2015.02.028
  [arXiv:1412.5525 [physics.ins-det]].

\bibitem{Kaspar:2016ofv} 
  J.~Kaspar {\it et al.},
  ``Design and performance of SiPM-based readout of PbF$_2$ crystals for high-rate, precision timing applications,''
  JINST {\bf 12}, no. 01, P01009 (2017)
  doi:10.1088/1748-0221/12/01/P01009
  [arXiv:1611.03180 [physics.ins-det]].

\bibitem{Midas}
MIDAS project, "MIDAS" [software], version 2.1, available from
https://bitbucket.org/tmidas/midas [accessed 2017-10-18]

\bibitem{Green:2012gv} 
  C.~Green, J.~Kowalkowski, M.~Paterno, M.~Fischler, L.~Garren and Q.~Lu,
  ``The Art Framework,''
  J.\ Phys.\ Conf.\ Ser.\  {\bf 396}, 022020 (2012).
  doi:10.1088/1742-6596/396/2/022020

\bibitem{ZeroMQ}
ZeroMQ project, "ZeroMQ" [software], version 4.2.2, Available from http://zeromq.org/intro:get-the-software [accessed 2017-10-18]

\bibitem{Nodejs}
Node.js  project, "Node.js" [software], version 6.11.4, Available from https://nodejs.org/en/download/ [accessed 2017-10-18]

\bibitem{plotly}
Plotly Technologies Inc., 
``Collaborative data science,''
Plotly Technologies Inc., Montr\"{e}al, QC, 2015, https://plot.ly

\bibitem{Brun:1997pa} 
  R.~Brun and F.~Rademakers,
 ``ROOT: An object oriented data analysis framework,''
  Nucl.\ Instrum.\ Meth.\ A {\bf 389}, 81 (1997).
  doi:10.1016/S0168-9002(97)00048-X

\bibitem{Beffart:2017gxr} 
  T.~Beffart, M.~Früh, C.~Haas, S.~Rajgopal, J.~Schwabe, C.~Wolff and M.~Szuba,
  ``RootJS: Node.js Bindings for ROOT 6,''
  arXiv:1704.07887 [cs.OH].

\bibitem{Box:2016uzj} 
  D.~Box {\it et al.},
  ``The FIFE Project at Fermilab : Computing for Experiments,''
  PoS ICHEP {\bf 2016}, 176 (2016).
 
 \bibitem{Eigen}
  Eigen project, “Eigen” [software], version 3.3.4, Available from https://bitbucket.org/eigen/eigen/ [accessed 2017-10-18]

\bibitem{pqxx}
pqxx project, "pqxx" [software], version 5.0.1, Available from https://github.com/jtv/libpqxx [accessed 2017-10-20]

\bibitem{Torjo13}
John Torjo,
``Boost.Asio C++ Network Programming,''
http://www.packtpub.com/boost-asio-cplusplus-network-programming/book

\bibitem{gm2ringsim} 
 gm2ringsim project, "gm2ringsim" [software], version 7.0.4, Available from https://cdcvs.fnal.gov/redmine/projects/gm2ringsim [accessed 2017-10-20]
  
\end{thebibliography}
\end{document}